\documentclass[twocolumn]{autart}
\usepackage{epsfig} 
\usepackage{amsmath} 
\usepackage{latexsym,amssymb}
\usepackage{ebezier}
\usepackage{graphicx}      
\usepackage{natbib}        
\usepackage{color}
\newtheorem{define}{Definition}
\newtheorem{remark}{Remark}

\newtheorem{lemma}{Lemma}

\begin{document}
\begin{frontmatter}

\title{Distributed Event-Triggered Control for Asymptotic Synchronization of Dynamical Networks\thanksref{footnoteinfo}}

\thanks[footnoteinfo]{The material in this paper was not presented at any conference. }

\author[First]{Tao Liu}\ead{taoliu@eee.hku.hk}
\author[Second]{Ming Cao}\ead{m.cao@rug.nl}
\author[Second]{Claudio De Persis}\ead{c.de.persis@rug.nl}
\author[Third]{Julien M. Hendrickx}\ead{ulien.hendrickx@uclouvain.be}

\address[First]{Department of Electrical and Electronic Engineering \\
The University of Hong Kong, Hong Kong S.A.R., China}
\address[Second]{Faculty of  Science and Engineering \\
University of Groningen, 9747 AG, Groningen, The Netherlands}
\address[Third]{ ICTEAM Institute, Universit\'e catholique de Louvain,  Louvain-la-Neuve, Belgium}

\begin{abstract}                
This paper studies synchronization of dynamical networks with event-based communication.  Firstly, two estimators are introduced into each node,  one  to estimate its own  state, and the other to estimate the average state of its neighbours. Then, with these  two estimators, a distributed event-triggering rule (ETR) with a dwell time is  designed such that the  network achieves  synchronization asymptotically with no Zeno behaviours. The designed ETR only depends on the information that each node can obtain, and thus can be implemented in a decentralized way.  

\end{abstract}


\end{frontmatter}
\section{Introduction}\label{sec_intro}

Synchronization of dynamical networks, and its related problem---consensus of multi-agent systems, have attracted a lot of attention due to their extensive applications in various fields (see  \cite{AArenas_phyRep_2008,ROlfati-Saber_ieeep_2007,Ren_csm07,CWWu_sync_2007} for details). Motivated by the fact that connected nodes in some real-world networks share information over a digital platform,  these problems have  recently been investigated under the circumstance that nodes  communicate to their neighbours only at certain discrete-time instants. To use the limited communication network resources effectively, event-triggered control (ETC) (see \cite{WHeemels_cdc_2012} and reference therein) introduced in  networked control systems  has been extensively used to synchronize networks. Under such a circumstance, each node can only get limited information, and the main issue becomes how to use these limited information to design an ETR for each node such that the network achieves synchronization asymptotically and meanwhile to prevent Zeno behaviours that are caused by the continuous/discrete-time hybrid nature of ETC, and undesirable in practice (\cite{PTabuada_tac_2007}). 

 Early works in ETC  focused on dynamical networks  with simple node dynamics  such as single-integrators and double-integrators. In \cite{DDimarogonas_cdc_2009}, distributed  ETC  was used to achieve consensus. To prevent  Zeno behaviour, a decentralized ETR with a time-varying threshold was introduced  to achieve  consensus  in \cite{GSeyboth_auto_2013}.  Self-triggered strategies were proposed in \cite{DePersis_tac_2013} and shown to be robust to  skews of the local clocks, delays, and limited precision in the communication.   

Most recently,  attention has been increasingly paid  to networks with generalized linear node dynamics. Different types of ETC have been developed to achieve either bounded or asymptotic synchronization for such networks (e.g., \cite{ODemir_adhs_2012, zhu_auto_2014, Tliu_necsys13, meng_auto,xiao_ijc, Garcia_acc_2015,Yang_auto_2016, Hu_cyb_2016}). In order to achieve asymptotic synchronization as well as to prevent Zeno behaviours, two main methods are developed in the literature. One uses bidirectional communication,  i.e., at each event time, a node  sends its sampled state to its neighbours and meanwhile asks for its neighbours' current states to update the control signal (e.g., \cite{meng_auto,xiao_ijc,Hu_cyb_2016}). The other uses unidirectional communication, i.e.,  a node only  sends its sampled information to its neighbours but does not require information from its neighbours (e.g., \cite{Tliu_necsys13,Garcia_acc_2015,Yang_auto_2016}). However, the latter needs  $d_i+1\geq 2$ estimators in each node and uses an exponential term in the ERT in oder to prevent Zeno behaviours. 

In this paper, we study asymptotic synchronization of networks with generalized linear node dynamics by using the unidirectional communication method. The main differences from the existing results are as follows. Firstly,  a new sampling mechanism is used with which two estimators are introduced into each node, whereas most existing results need every node to estimate the states of all its neighbours.   Secondly,  inspired by the method proposed in  \cite{Talla_tac_2014},  we replace the exponential term  extensively used in the literature by a dwell time that was originally introduced in  switched systems (\cite{MCao_scl2010}), which can simplify the implementation of the designed ETR. Thirdly, a distributed ETR for each node is designed based on the two estimators and dwell time, whereas most of the existing results use decentralized ETRs that only consist of local information of the node itself, i.e., the state error between the node and its own estimator and the time-dependent exponential term (e.g., \cite{Garcia_acc_2015,Yang_auto_2016}). By introducing an estimation of the synchronization errors between neighbours using the neighbours' sampled information, the proposed ETR method can reduce the number of sampling times for each node significantly. 

\section{Network Model and Preliminaries}\label{sec_model}

\emph{Notation}:  Denote  the set of real numbers, non-negative real numbers,  and  non-negative integers by $\mathbb{R}$, $\mathbb{R^+}$, and $\mathbb{Z}^+$; the set of $n$-dimensional real vectors and $n\times m$ real matrices by $\mathbb{R}^n$ and $\mathbb R^{n\times m}$. $I_n$, $1_n$ and $1_{n\times m}$ are the $n$-dimensional identity matrix,  $n$-dimensional vector and  $n\times m$ matrix with all entries being $1$, respectively. 
$\|\cdot\|$ represents the Euclidean norm for vectors and also the induced norm for matrices. The superscript $(\cdot)^\top$ is the transpose of  vectors or matrices.  $\otimes$ is the Kronecker product of matrices. For a single $\omega:~\mathbb{R^+}\to\mathbb{R}^n$, $\omega(t^-)=\lim_{s\uparrow t}\omega(s)$.
Let   $\mathcal{G}$ be an undirected graph consisting of  a node set  $\mathcal{V}=\{1,2,\dots, N\}$ and a link set $\mathcal{E}=\{\bar e_1,\bar e_2,\dots, \bar e_M\}$.  If there is a link $\bar e_k$  between nodes $i$ and $j$, then we say node  $j$ is a neighbour of node $i$ and vice versa. Let $A=(a_{ij})\in \mathbb{R}^{N\times N}$ be the adjacency matrix of $\mathcal{G}$, where  $a_{ii}=0$ and $a_{ij}=a_{ji}>0$, $i\neq j$, if node $i$ and node $j$ are neighbours, otherwise $a_{ij}=a_{ji}=0$. The Laplacian matrix $L=(l_{ij})\in\mathbb{R}^{N\times N}$ is defined by $l_{ij}=-a_{ij}$, if $j\neq i$ and $l_{ii}=\sum_{j=1}^Na_{ij}$.
 
We consider a  dynamical network described by
\begin{equation}\label{nm_obb}
\dot x_i(t)=Hx_i(t)+B u_i(t),~~\forall i\in\mathcal{V}
\end{equation}
where  $x_{i}=(x_{i1},x_{i2},\dots,x_{in})^\top\in{\mathbb{R}^{n}}$
is the state of node $i$.  $H\in\mathbb{R}^{n\times n}$, $B\in \mathbb{R}^{n}$, and $u_i\in{\mathbb{R}}$ are the  node dynamics matrix,  input matrix, and control input, respectively. Generally, continuous communication between neighbouring nodes is assumed, i.e., $u_i(t)=K\sum_{j=1}^Na_{ij}(x_j(t)-x_i(t))$. This yields the following network 
\begin{equation}\label{continuous_m}
\dot x_i(t)=Hx_i+BK\sum\nolimits_{j=1}^Na_{ij}(x_j(t)-x_i(t)).
\end{equation}
In this paper,  we assume that connections in \eqref{nm_obb}  are realized via discrete communication, i.e., each node only obtains information from its neighbours  at certain discrete-time instants.  We will present an event-triggered version of network \eqref{continuous_m}, and study how to design an ETR for each node to achieve asymptotic synchronization.   We suppose that the topological structure of the network is  fixed, undirected and connected.   

We introduce two estimators $\mathcal O_i$ and $\mathcal O_{\mathcal{V}_i}$ into each node $i$, where $\mathcal O_i$ is used  to estimate its own state, and  $\mathcal O_{\mathcal{V}_i}$ is used to estimate the average state of its neighbours. We adopt the  following control input
\begin{align}\label{coupling_2a}
u_i(t)=K\left(\hat x_{\mathcal{V}_i}(t)-l_{ii}\hat x_i(t)\right)
\end{align}
where $K\in\mathbb{R}^{1\times n}$ is the control gain to be designed, $\hat x_i\in\mathbb{R}^n$ and $\hat x_{\mathcal{V}_i}\in\mathbb{R}^n$ are states of $\mathcal O_i$ and $\mathcal O_{\mathcal{V}_i}$, respectively. The state equations of $\mathcal O_i$ and $\mathcal O_{\mathcal{V}_i}$ are given by
\begin{align}\label{estimators}
\mathcal O_i:~&\begin{array}{ll}
\dot {\hat x}_i(t)=H\hat x_i(t), & \hbox{$t\in[t_{k_i},t_{{k_i}+1})$}\\
\hat x_i(t)=x_i(t), & \hbox{$t=t_{k_i}$}\\
\end{array}
\end{align}\vspace{-15mm}

\begin{align}\label{estimator_b}
\mathcal O_{\mathcal{V}_i}: ~&\begin{array}{ll}
\dot {\hat x}_{\mathcal{V}_i}(t)=H\hat x_{\mathcal{V}_i}(t), ~~~~~~~~~~~~~t\in[t_{\bar k_i},t_{{\bar k_i}+1})\\
\hat x_{\mathcal{V}_i}(t)=\hat x_{\mathcal{V}_i}(t^-)-\!\sum\nolimits_{j\in\mathcal{J}_i}e_j(t^-),~~~ t=t_{\bar k_i}.
\end{array}
\end{align}
The increasing time sequences $\{t_{k_i}\}$ and $\{t_{\bar k_i}\}$, $k_i, ~\bar k_i \in\mathbb{Z}^+$ represent time instants that node $i$ sends updates to its neighbours and that it receives updates from one or more of its neighbours, respectively.  We assume that: there is no time delay for computation and execution, i.e., $t_{k_i}$ represents both the $k_i$th sampling time  and the $k_i$th time when node $i$ broadcasts updates;  and the communication network is under an ideal circumstance, i.e.,  there are no time delays or data dropouts in communication.  Therefore, the set ${\mathcal{J}_i}={\mathcal{J}_i}(t_{\bar k_i})=\left\{j~|~t_{k_j}=t_{\bar k_i}, j\in\mathcal{V}_i\right\}$ is a subset of $\mathcal{V}_i$, from which node $i$ receives updated information at $t=t_{\bar k_i}$, and $\mathcal{V}_i=\{j~|~a_{ij}>0,~j\in\mathcal{V}\}$ is the index set of the neighbours for node $i$.   The vector $e_i(t)=\hat x_i(t)-x_i(t)$  represents the  deviation between the state of estimator $\mathcal O_i$  and its own, and which node $i$ can easily compute.  

The time sequence $\{t_{k_i}\}$ is decided by the   ETR  
\begin{equation}\label{etc_ga}
t_{k_i+1}=\inf\left\{t> t_{k_i}~|~r_i(t,x_i,\hat x_i,\hat x_{\mathcal{V}_i})>0\right\}
\end{equation}
where   $r_i(\cdot,\cdot,\cdot,\cdot):\mathbb{R}^+\times\mathbb{R}^n\times\mathbb{R}^n\times\mathbb{R}^n\to\mathbb{R}$ is the event-triggering function to be designed.
For $t>t_{k_i}$, if $r_i>0$ at $t=t_{k_i+1}^-$, then node $i$ samples $x_i(t_{k_i+1}^-)$, $\hat x_i(t_{k_i+1}^-)$,  calculates $e_i(t_{k_i+1}^-)$, sends $e_i(t_{k_i+1}^-)$ to its neighbours, and reinitialize the estimator $\mathcal O_i$ at $t=t_{k_i+1}$ by  $x_i(t_{k_i+1})$. In addition, node $i$ will reinitialize the estimator $\mathcal O_{\mathcal{V}_i}$ by $\hat x_{\mathcal{V}_i}(t_{\bar k_i+1})=\hat x_{\mathcal{V}_i}(t_{\bar k_i+1}^-)-\sum_{j\in\mathcal{J}_i}e_j(t_{\bar k_i+1}^-)$ each time when it receives updates from its neighbours. We further assume  the network is well initialized at $t=t_0$, i.e., $\hat x_i(t_0)=0$ and each node samples and sends $e_i(t_0)$ to its neighbours. Therefore, we have $\hat x_i(t_0)=x_i(t_0)$, $\hat x_{\mathcal{V}_i}(t_0)=\sum_{j\in\mathcal{V}_i}x_j(t_0)$ and $\mathcal{J}_i(t_0)=\mathcal{V}_i$ for all $i\in\mathcal{V}$.  Then, the problem is with the given network topology, to  design a proper ETR \eqref{etc_ga} such that network \eqref{nm_obb} achieves synchronization asymptotically without Zeno behaviours.

To simplify the analysis, we will show that network \eqref{nm_obb} with controller \eqref{coupling_2a} and estimators \eqref{estimators}, \eqref{estimator_b} is equivalent to the following system where each node maintains an estimator of the state of each of its neighbours. 
\begin{subequations}\label{nm_ob}
\begin{align}
&\dot x_i(t)=Hx_i(t)-BK\sum\nolimits_{j=1}^Nl_{ij} \hat x_j(t),\forall i\in\mathcal{V}\label{nm_ob1}\\
\mathcal{O}_i: &\begin{array}{ll}
\dot {\hat x}_i(t)=H\hat x_i(t), & \hbox{$t\in[t_{k_i},t_{{k_i}+1})$}\\
\hat x_i(t)=x_i(t), & \hbox{$t=t_{k_i}$.}\\
\end{array}\label{nm_ob2}
\end{align}
\end{subequations}
Defining $\bar z_i=\sum_{j\in\mathcal{V}_i}\hat x_j$ gives
$\dot{\bar z}_i(t)=\sum_{j\in\mathcal{V}_i}\dot{\hat x}_j(t)=H\bar z_i(t)$, $t\in[t_{\bar k_i},t_{\bar k_i+1})$,
which has the same dynamics as $\hat x_{\mathcal{V}_i}$ defined in \eqref{estimator_b}. Moreover, at $t=t_{\bar k_i}$, we have 
\begin{align}
\bar z_i(t)&=\sum_{j\in\mathcal{V}_i/\mathcal{J}_i(t)}\hat x_j(t^-)+\sum_{j\in\mathcal{J}_i(t)}x_j(t)\notag\\
&=\sum_{j\in\mathcal{V}_i/\mathcal{J}_i(t)}\hat x_j(t^-)+\sum_{j\in\mathcal{J}_i(t)}\left(\hat x_j(t^-)-e_j(t^-)\right)\notag\\
&=\hat x_{\mathcal{V}_i}(t).
\end{align}
Thus, we have $\bar z_i(t)=\hat x_{\mathcal{V}_i}(t)$ for all $t\geq t_0$. Then, controller \eqref{coupling_2a} becomes 
\begin{equation}\label{equ_ab}
u_i=K\left(\bar z_i-l_{ii}\hat x_i\right)=K\left(\hat x_{\mathcal{V}_i}-l_{ii}\hat x_i\right).
\end{equation}
Substituting \eqref{equ_ab} into \eqref{nm_obb} gives that  network \eqref{nm_obb} with \eqref{coupling_2a}, \eqref{estimators}, and \eqref{estimator_b} is equivalent to \eqref{nm_ob}.  

Moreover, let $\hat z_i=\sum_{j\in\mathcal{V}_i}(\hat x_j-\hat x_i)$. We have $\hat x_{\mathcal{V}_i}=\bar z_i=\hat z_i+l_{ii}\hat x_i$. Then, ETR \eqref{etc_ga} can be reformulated as
\begin{equation}\label{etc_g}
t_{k_i+1}=\inf\left\{t> t_{k_i}~|~r_i(t,x_i,\hat x_i,\hat z_i)>0\right\}.
\end{equation}
In network \eqref{nm_ob},  $\hat z_i$ in ETR \eqref{etc_g} contains information of $\hat x_j$, $j\in\mathcal{V}_i$ which are not available for node $i$ as node $i$ only has  estimator \eqref{nm_ob2}. Therefore,  one estimator for each node is insufficient to implement ETR \eqref{etc_g} in practice. To overcome this difficulty, we introduce another estimator \eqref{estimator_b} into each node. It is shown that network  \eqref{nm_ob} is theoretically equivalent to network \eqref{nm_obb} with the two estimators $\mathcal{O}_i$, $\mathcal{O}_{\mathcal{V}_i}$,  and ETR \eqref{etc_g}  is equivalent to ETR \eqref{etc_ga} which can be implemented in practice.

\begin{remark}
It is shown in \cite{Tliu_necsys13} that under the same assumptions, a network with $d_i+1$ estimators for each node ($d_i$ is the number of neighbours of node $i$) is also theoretically equivalent to network  \eqref{nm_ob}, and thus equivalent to network \eqref{nm_obb} with two estimators $\mathcal{O}_i$ and $\mathcal{O}_{\mathcal{V}_i}$. On the other hand, the error $e_i(t)=\hat x_i(t)-x_i(t)$  is extensively used in the literature to design ETR, where each node  sends its sampled state to its neighbours.  By having each node sending $e_i(t_{k_i})$ instead of $x_i(t_{k_i})$, it turns out that we can reduce the number of estimators.  The implementation of this new sampling mechanism needs no more information than that used in the literature. Further,  instead of calculating $d_i+1\geq 2$ estimators $\hat x_j$,  our method only calculates $\hat x_i$ and $\hat x_{\mathcal{V}_i}$ for each node $i$, and hence, our method has implementation advantages, in particular for networks with large $d_i$ and limited embedded computing resources in each node. Like most of the existing results in the  literature of ETC, in our method each node needs to send $e_i(t)$ (or $x_i(t)$) to its neighbours rather than the relative state information $(x_j(t)-x_i(t))$ that is extensively used in network \eqref{continuous_m} with continuously interconnected nodes. Of course, it is important to study network \eqref{nm_ob}  by only using the relative state information for the design purposes which should be studied in the future. 
\end{remark}

This paper will use  model \eqref{nm_ob} and  ETR \eqref{etc_g} for the analysis. But the obtained results can be implemented by using controller \eqref{coupling_2a} with the two estimators $\mathcal{O}_i$, $\mathcal{O}_{\mathcal{V}_i}$ and ETR \eqref{etc_ga}.  Based on network \eqref{nm_ob}, we give the definition of asymptotic synchronisation.
\begin{define}\label{sync}
Let $x(t)=\left(x_1^\top(t),x_2^\top(t),\dots,x_N^\top(t)\right)^\top \in\mathbb{R}^{nN}$ and $\hat x(t)=\left(\hat x_1^\top(t),\hat x_2^\top(t),\dots,\hat x_N^\top(t)\right)^\top \in\mathbb{R}^{nN}$ be a  solution of  network \eqref{nm_ob} with  initial condition $x_{0}=(x_{10}^\top, x_{20}^\top,\dots,$ $x_{N0}^\top)^\top$ and $x_{i0}=x_i(t_0)$. Then, the network is said to achieve \emph{synchronization asymptotically}, if  for every  $x_{0}\in\mathbb{R}^{nN}$ the following condition is satisfied
\begin{equation}\label{ed}
\lim_{t\to\infty}\|x_i(t)-x_j(t)\|=0,~~\forall~i,j\in\mathcal{V}.
\end{equation}
\end{define}
\begin{remark}
When the communication network is not ideal, model  \eqref{nm_obb} with \eqref{coupling_2a} and $\mathcal{O}_i$, $\mathcal{O}_{\mathcal{V}_i}$ cannot be simplified to \eqref{nm_ob}.   A more  complicated model is needed to describe the network dynamics. Time delays and packet loss will influence the synchronization performance. However, due to the robust property of asymptotic synchronization, bounded synchronization can be guaranteed where the final synchronization error may depend on the time delay magnitude and probability of packet loss.  Another important problem for this case is under what conditions  the network can still  achieve synchronization asymptotically. These issues should be studied in the future.   
\end{remark}
\section{Event-Triggered Control}\label{sec_sync}

Denote  $e(t)=\left(e_1^\top(t),e_2^\top(t),\dots,e_N^\top(t)\right)^\top$ with $e_i(t)=\hat x_i(t)-x_i(t)$. Network \eqref{nm_ob1} can be rewritten by\begin{equation}\label{nm_full}
\dot x=(I_N\otimes H-L\otimes BK)x-(L\otimes BK) e.
\end{equation}
Since the topology of the network is undirected and connected,  the Laplacian matrix $L $ is irreducible, symmetric, and  has only  one zero eigenvalue. Further,  there exists an orthogonal  matrix $\Psi=(\psi_1,\psi_2,\dots,\psi_N)\in\mathbb{R}^{N\times N}$ with $\psi_i=(\psi_{i1},\psi_{i2},\dots,\psi_{iN})^\top$ and $\Psi^\top\Psi=I_N$ such that
$\Psi^\top L\Psi=\Lambda=\textup{diag}(\lambda_1,\lambda_2,\dots,\lambda_N)$
where $0=\lambda_1<\lambda_2\leq\lambda_3\leq\dots\leq\lambda_N$. Choose $\psi_1={1}/{\sqrt{N}}1_N^\top$ for $\lambda_1$. Due to the zero row sum property of $L$, we have   $\sum_{j=1}^N\psi_{ij}=0$ for all $i=2,3,\dots,N$. Defining $\Phi=(\psi_2,\psi_3,\dots,\psi_N)\in\mathbb{R}^{N\times(N-1)}$ gives 
\begin{equation}\label{PHI_pro}
\begin{split}
\Phi^\top\Phi=I_{N-1},~~~~\Phi\Phi^\top=I_N-\frac{1}{N}1_{N\times N}.
\end{split}
\end{equation}
Let $\Lambda_1=\Phi^\top L\Phi=\textup{diag}\{\lambda_2,\lambda_3,\dots,\lambda_N\}$, $\bar\Phi=\Phi\otimes I_n$ and $\bar\Lambda=\Lambda_1\otimes BK=\textup{diag}\left\{\lambda_2BK,\lambda_3BK,\dots,\lambda_NBK\right\}$. Defining $y=\bar\Phi^\top x$ gives
\begin{align}\label{nm_y1}
\dot y(t)=&\bar\Phi^\top\left((I_N\otimes H)x-(L\otimes  BK)(I_{Nn}-\bar\Phi\bar\Phi^\top\right.\notag\\
&\left.+\bar\Phi\bar\Phi^\top)(x+ e)\right)\notag\\
=&(I_{N-1}\otimes H-\Lambda_1\otimes BK)y-\bar\Lambda\bar\Phi^\top e
\end{align}
where we use properties $\bar\Phi^\top(I_N\otimes H)=(I_{N-1}\otimes H)\bar\Phi^\top$ and $(L\otimes  BK)(I_{Nn}-\bar\Phi\bar\Phi^\top)=0$ for any $ BK$, which are supported by facts $L1_N=0$ and \eqref{PHI_pro}. Denoting  $\bar H=(I_{N-1}\otimes H)-(\Lambda_1\otimes BK)=\textup{diag}\left\{H_2,H_3,\dots,H_N\right\}$ with $H_i=H-\lambda_iBK$, system \eqref{nm_y1} can be simplified to
\begin{equation}\label{nm_y}
\dot y=\bar Hy-\bar\Lambda\bar\Phi^\top e.
\end{equation}
By defining $\bar x=\frac{1}{N}\sum_{i=1}^Nx_i$, we have $\|y\|^2=x^\top\bar\Phi\bar\Phi^\top x
=\sum_{i=1}^N\|x_i-\bar x\|^2$ where the last equality  follows from  $\Phi^\top\Phi=I_{N-1}$ and $(\bar\Phi\bar\Phi^\top)^2=\bar\Phi\bar\Phi^\top$. Therefore, if $\lim_{t\to\infty}\|y(t)\|=0$, then  $x_i(t)$, $x_j(t)$, and $\bar x(t)$ are asymptotically equal when  $t\to \infty$, i.e., network \eqref{nm_ob} achieves synchronization asymptotically. This result is summarized in the following lemma.

\begin{lemma}\label{lem2}
If system \eqref{nm_y}  is asymptotically stable, i.e., $\lim_{t\to\infty}\|y(t)\|=0$,
then network \eqref{nm_ob} achieves synchronization asymptotically.
\end{lemma}

It is shown in \cite{HTrentelman_tac_2013} that  a necessary and sufficient condition for asymptotic synchronization of network  \eqref{continuous_m} with continuous interconnections  is the existence of positive definite matrices $P_i$ such that
\begin{equation}\label{cetle}
H_i^\top P_i+P_iH_i=-2I_n,~i=2,3,\dots,N.
\end{equation}
This condition requires all the linear systems with system matrices $H_i=H-\lambda_iBK$, $i=2,\dots,N$ are asymptotically stable simultaneously, which is stronger than that $(H,B)$ is stabilizable. Another alternative is to find a common $P>0$ for all $H_i$, $i=2,\dots,N$ (e.g., \cite{Wu_auto_2017}).  From \eqref{nm_y1}, network \eqref{nm_ob} with ETC can be regarded  as network \eqref{continuous_m} with an external input (or a disturbance) $\bar\Lambda\bar\Phi^\top e$. According to input-to-state stability theory, a necessary condition for  system  \eqref{nm_y1} to be asymptotically stable  is that the the corresponding system (also described by \eqref{nm_y1}  but without the term $\bar\Lambda\bar\Phi^\top e$) is asymptotically stable. Hence, the existence of matrix solutions $P_i$ to Lyapunov equations  \eqref{cetle} is also a fundamental requirement for network  \eqref{nm_ob} with ETC to achieve asymptotic synchronization. In this paper, we  assume that  such matrices  $P_i$ exist.  

Let $z_i=\sum_{j\in\mathcal{V}_i}(x_j-x_i)$, $\hat z_i=\sum_{j\in\mathcal{V}_i}(\hat x_j-\hat x_i)$,  $z=(z_1^\top,z_2^\top,\dots,z_N^\top)^\top=(-L\otimes I_n)x$, and $\hat z=(\hat z_1^\top,\hat z_2^\top,\dots,\hat z_N^\top)^\top=(-L\otimes I_n)\hat x$. Next, we give a useful lemma which will be used to prove the main result. 

\begin{lemma}\label{lemma0}
Consider network \eqref{nm_ob}. The following  two inequalities hold for any $t\geq t_0$
\begin{align}\label{norm_z}
\|\hat z\|&\leq\lambda_N(\|e\|+\|y\|)\\
\lambda_2\|y\|&\leq\lambda_N\|e\|+\|\hat z\|.\label{norm_y}
\end{align}
\end{lemma}

\begin{pf} 
Due to $\|(L\otimes I_n)\|=\lambda_N$, we have
\begin{equation}\label{z_hat}
\|\hat z\|=\|(L\otimes I_n)(x+e)\|\leq\|z\|+\lambda_N\|e\|
\end{equation}
\begin{equation}\label{zn}
\|z\|=\|(L\otimes I_n)(\hat x-e)\|\leq\|\hat z\|+\lambda_N\|e\|.
\end{equation}
Let $U=\Phi\Phi^\top$, then  for any $L$, we have $LU=UL$, i.e., $L$ and $U$ are diagonalizable simultaneously. Further, we have
$\Psi^\top L\Psi=\Lambda$ and $\Psi^\top U\Psi=\textup{diag}\{\lambda_{u1},\lambda_{u2},\dots,\lambda_{uN}\}$, where $\lambda_{u1}=0$ and $\lambda_{ui}=1$, $i=2,3,\dots,N$ are eigenvalues of $U$. Let $\bar\lambda_i$, $i=1,2,\dots,N$ be  eigenvalues of the matrix $(\lambda_N^2U^2-L^2)$. Then with $U^2=U$, we have $\bar\lambda_1=0$ and $\bar\lambda_i=\lambda_N^2-\lambda_i^2\geq0$, $i=2,3,\dots,N$, which gives $L^2\leq\lambda_N^2U^2$. Thus, we have
\begin{equation}\label{a2}
\begin{split}
\|z\|^2=&x^\top(L^2\otimes I_n)x\leq\lambda_N^2x^\top(U^2\otimes I_n)x\\
=&\lambda_N^2\|\bar\Phi^\top x\|^2=\lambda_N^2\|y\|^2.
\end{split}
\end{equation}
Combining \eqref{z_hat} with \eqref{a2} gives  inequality \eqref{norm_z}. Similar to \eqref{a2}, we have $\|y\|^2=x^\top(U^2\otimes I_n)x\leq1/\lambda_2^2x^\top(L^2\otimes I_n)x$ which with  \eqref{zn} gives \eqref{norm_y}. \hfill$\Box$
\end{pf}
Let $\rho=\frac{\delta}{\lambda_N\sqrt{2N(\alpha^2+\delta^2)}}$, $\rho_1=\frac{1}{\lambda_2}(\frac{\delta}{\sqrt{2(\alpha^2+\delta^2)}}+1)$, $\delta\in(0,1)$, $\alpha=\max_{i=2,3,\dots,N}$ $\{\lambda_i\|P_i BK\|\}$, $a=\|H\|+\|\bar H\|+\lambda_N\frac{\delta}{\alpha}\| BK\|$,  $b=\lambda_N\| BK\|(1+\frac{\delta}{\alpha})$, and 
$\tau^*=\frac{1}{a}\ln\left(\frac{a\rho}{b\rho_1}+1\right)>0$.
We have the following result.

\begin{thm}\label{them5}
Network \eqref{nm_ob}  achieves synchronization asymptotically under the distributed ETR
\begin{equation}\label{detre_dwell}
t_{k_i+1}=\inf\left\{t\geq t_{k_i}+\tau^*~|~\|e_i\|>\rho\|\hat z_i\|\right\}.
\end{equation}
 Moreover, no Zeno behaviour occurs in the network.
\end{thm}
\begin{pf}
Under ETR \eqref{detre_dwell}, the existence of $\tau_{k_i}=t_{k_i+1}-t_{k_i}>0$  is guaranteed by the dwell time $\tau^*$.  To show  asymptotic synchronization, we claim that the network with \eqref{detre_dwell} satisfies 
\begin{equation}\label{e-hat_z}
\|e_i\|\leq\rho\|\hat z\|, ~~\forall i\in\mathcal{V},~\forall t\geq t_0.
\end{equation}
This is true at $t=t_0$, as we have $\|e_i(t_0)\|=0$ and hence $\|e_i(t_0)\|\leq\rho\|\hat z(t_0)\|$, $\forall i\in\mathcal{V}$.  Suppose to obtain a contradiction that \eqref{e-hat_z} does not always hold, and let $t^*$ be the infimum of times at which it does not hold. Due to the finite number of nodes, there exists a node $l$ such that $\|e_l\| > \rho \|\hat z\|$ for times arbitrarily close  $t^*$ from above, i.e., $\forall\epsilon>0$, $\exists t \in [t^*, t^*+\epsilon]$ such that $\|e_l(t)\| > \rho \|\hat z(t)\|$. It follows from ETR \eqref{detre_dwell} that $t^*$ must be in $(t_{k_l}, t_{k_l} + \tau^*]$ for some $k_l\in\mathbb{Z}^+$. We now show that there cannot exist a $t^*$ in $(t_{k_l}, t_{k_l} + \tau^*]$, which will establish \eqref{e-hat_z}. Since $\|e_i(t)\|\leq\rho\|\hat z(t)\|$,  $\forall i\in\mathcal{V}$, $\forall t<t^*$, which gives
\begin{equation}\label{e_norm1}
\|e\|^2=\sum_{i=1}^N\|e_i\|^2\leq\frac{\delta^2}{2\lambda_N^2(\alpha^2+\delta^2)}\|\hat z\|^2.
\end{equation}
On the other hand, inequality \eqref{norm_z} gives 
\begin{equation}\label{norm_zsqrt}
\|\hat z\|^2\leq2\lambda_N^2(\|e\|^2+\|y\|^2).
\end{equation}
Substituting \eqref{norm_zsqrt} into \eqref{e_norm1} yields 
\begin{equation}\label{norm_e1}
\|e(t)\|\leq\frac{\delta}{\alpha}\|y(t)\|, ~~\forall t\in[t_0,t^*).
\end{equation}
Calculating $\frac{d}{dt}\frac{\|e_l\|}{\|y\|}$ for $t\in[t_{k_l},t^*)$ directly gives
\begin{align}\label{dot_1}
\frac{d}{dt}\frac{\|e_l\|}{\|y\|}
\leq&\left(\|H\|+\|\bar H\|\right)\frac{\|e_l\|}{\|y\|}+\frac{\|\bar\Lambda\|\|e_l\|\|e\|}{\|y\|^2}\notag\\
&+\lambda_N\| BK\|\frac{\|e\|}{\|y\|}+\lambda_N\|\ BK\|
\end{align}
where we use \eqref{norm_z} in Lemma \ref{lemma0} to get \eqref{dot_1}.  Substituting \eqref{norm_e1} into \eqref{dot_1} gives 
\begin{equation}\label{ei/y_dot}
\frac{d}{dt}\frac{\|e_l\|}{\|y\|}\leq a\frac{\|e_l\|}{\|y\|}+b.
\end{equation}
Based on the comparison theory (\cite{Khalil_2002}), we have
$\|e_l(t)\|/\|y(t)\|\leq\phi(t-t_{k_l})$,  whenever  $\|e_l(t_{k_l})\|$ $/\|y(t_{k_l})\|\leq\phi(t_{k_l})$
where  $\phi(t-t_{k_l})$ is the solution of the ordinary differential equation 
\begin{equation}\label{tau_equ2}
\dot \phi=a\phi+b
\end{equation}
with the initial condition $\phi(t_{k_l})$.   At $t=t_{k_l}$, we have $\|e_l(t_{k_l})\|/\|y(t_{k_l})\|=0$. Setting $\phi(t_{k_l})=0$ gives 
\begin{equation}\label{e-phi}
\frac{\|e_l(t)\|}{\|y(t)\|}\leq\phi(t-t_{k_l}), \forall t\in [t_{k_l}, t^*).
\end{equation}
Further, combining \eqref{norm_y} with \eqref{e_norm1} gives $\|\hat z\|\geq\|y\|/\rho_1$ which with \eqref{e-phi} leads to
$$\frac{\|e_l(t)\|}{\|\hat z(t)\|}\leq\rho_1\frac{\|e_l(t)\|}{\|y(t)\|}\leq\rho_1\phi(t-t_{k_l}), ~~\forall t\in[t_{k_l},t^*).$$
Solving \eqref{tau_equ2} with $\phi(t_{k_l})=0$ shows that it will take $\phi(t-t_{k_l})$ a positive time constant $\tau^*$ to change its values from 0 to $\rho/\rho_1$, so does $\|e_l(t)\|/\|y(t)\|$. Therefore, it requires at least $\tau^*$ to make  $\|e_l(t)\|$ move from 0 to $\rho\|\hat z(t)\|$. 

Suppose, to obtain a contradiction, that $t^* < t_{k_l} + \tau^*$. In that case, $\|e_l(t)\|/\|y(t)\| \leq \phi(t-t_{k_l}) < \phi(\tau^*) \leq \rho/\rho_1$, for all $t\leq t^*$.   By continuity of $\|e_l\|/ \|y\|$, this implies the existence of an $\varepsilon>0$ such that $\|e_l(t)\|/\|y(t)\| <\phi(\tau^*)$ for all $t \leq t^*+\varepsilon$. Therefore, there holds then $\|e_l(t)\| < \rho  \|\hat z(t)\|$ for all $t < t^*+ \varepsilon$, in contradiction with $t^*$ being the infimum of times at which $\|e_l(t)\| > \rho\|\hat z(t)\|$.

Now,  select the Lyapunov function candidate $V=y^\top Py$ with $P=\text{diag}\{P_2,P_3,\dots,P_N\}$. Then, the derivative of $V$ along  system \eqref{nm_y} satisfies
\begin{equation}\label{lyap_dot}
\dot V\leq-2\|y\|^2+2\alpha\|y\| \|\bar\Phi^\top e\|.
\end{equation}
The inequality \eqref{e-hat_z} holds, so does \eqref{norm_e1}. Combining \eqref{norm_e1} with $\|\bar\Phi\|=1$ yields 
\begin{equation}\label{phi_norma}
\|\bar\Phi^\top e\|\leq\|\bar\Phi^\top\|\|e\|=\|e\|\leq\frac{\delta}{\alpha}\|y\|.
\end{equation}
Substituting \eqref{phi_norma} into \eqref{lyap_dot} gives
\begin{equation}\label{lyap_dota}
\dot V\leq-2(1-\delta)\|y\|^2<0,~ \forall \|y\|\neq0.
\end{equation}
Therefore, the equilibrium point $y=0$ of system \eqref{nm_y} is asymptotically stable. Based on Lemma \ref{lem2}, the network achieves synchronization asymptotically.
\hfill$\Box$
\end{pf}

\begin{remark}
Like most  results in synchronization of dynamical network with/without ETC (e.g.,  \cite{HTrentelman_tac_2013, Guinaldo_2013_IET}), one usually needs some global parameters to guarantee asymptotic synchronization and exclude Zeno behaviours. These  parameters can be estimated  by using methods proposed in the related literature (e.g., \cite{Franceschelli_auto_2013}), and can be initialized to each node at the beginning.  However,  how to use local parameters rather than global ones (e.g., how to replace $N$ by using local parameter such as the degree of the node $d_i$)  remains open, and deserves attention.  
\end{remark}

\begin{remark}\label{remak_z}
Most of the existing results (e.g., \cite{ODemir_adhs_2012, Guinaldo_2013_IET, GSeyboth_auto_2013, zhu_auto_2014,Garcia_acc_2015,Yang_auto_2016} ) use decentralized ETRs which can be summarized in the following compact from 
\begin{equation}\label{etr_det}
t_{k_i+1}=\inf\left\{t~|~\|e_i\|>c_0+c_1\text{e}^{-\gamma t}\right\}
\end{equation}
where  $c_0\geq0$, $c_1\geq0$, $\gamma>0$ are three design parameters. Obviously, ETR \eqref{etr_det} only depends on local information from node $i$ itself, and  asymptotic synchronization can be achieved only when $c_0=0$.  In our paper,  we  introduce $\|\hat z_i\|$ into the proposed ETR  \eqref{detre_dwell}. The term $\|\hat z_i\|$  updated by $x_j(t_{k_j})$ estimates the synchronization errors between neighbours continuously, and thus provides each node useful information for determining its sampling times. Therefore, the proposed ETR can reduce the sampling times significantly, in particular for cases where  $\|\hat z_i\|$ is large (see the example in Section 4 for details). Further, it is shown in \cite{Tliu_necsys13} that a similar distributed ETR as \eqref{detre_dwell} but with an exponential term $c_1\text{e}^{-\gamma t}$ can also guarantee asymptotic synchronization. However, this paper replaces the exponential term  by a dwell time  which can be implemented easily in practice. Such a $\tau^*$ gives an upper bound for the designed ETR \eqref{detre_dwell}, and therefore, a modified ETR with $0<\tau_i^*\leq\tau^*$ can also synchronize the network without Zeno behaviours.
\end{remark}

\begin{remark}\label{rem_last}
 To simplify notations, this paper only considers the case where $u_i$ is a scalar. However, the obtained results can extend to multiple-input case directly. It is pointed out in \cite{heemels_tac} that the joint design of the controller and event-triggering rule is a hard problem. However, we can select any control gain $K$ to synchronizes the continuous-time network  \eqref{continuous_m}, i.e., to stabilize $(H, \lambda_i B)$, $i=2,\dots,N$ simultaneously. It can be selected by solving  a group of linear matrix inequalities. Moreover,  a periodic ETC method was proposed to stabilize linear systems in \cite{heemels_tac} where the triggering condition was verified periodically.  In the paper, we do not check the event-triggering condition in the time interval $[t_{k_i}, t_{k_i}+\tau^*)$, but  check the condition continuously during the period $[t_{k_i}+\tau^*, t_{k_i+1})$.  It is of great interest to study asymptotic synchronization by using periodic ETC and one-directional communications.
\end{remark}

\section{An Example}\label{sec_exam}

\begin{figure}[h]
\begin{center}
\includegraphics[width=8.5cm, height=6cm]{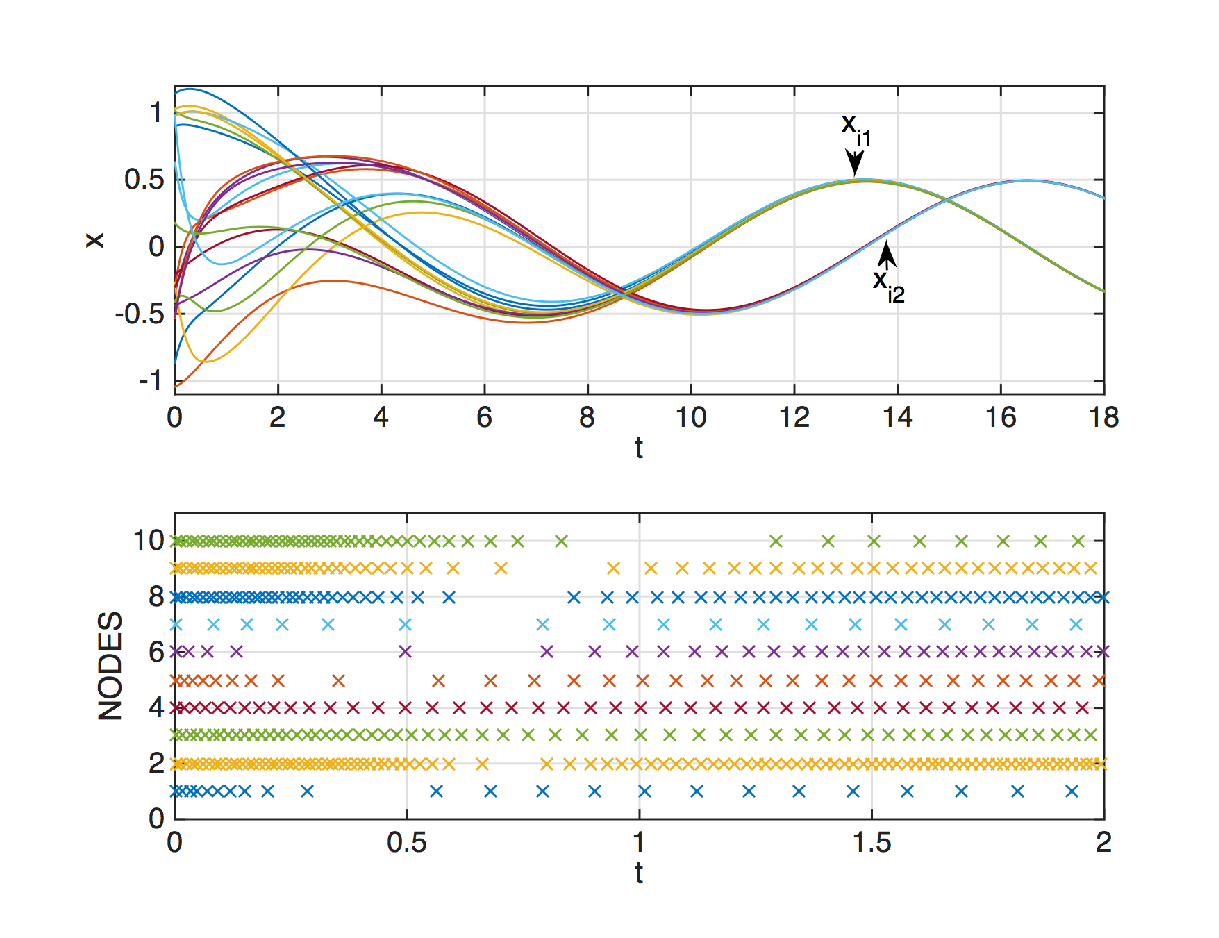} 
\caption{Simulation for the network with DDT.}
\label{fig_ddt}
\end{center}
\end{figure}

To show the effectiveness of our method,  consider a network with 10 nodes  that have  parameters as follows
\begin{equation*}
H=\left(%
\begin{array}{ccc}
  0 & -0.5  \\
  0.5 & 0 \\
\end{array}%
\right),~~B=\left(%
\begin{array}{c}
  0  \\
  1  \\
\end{array}%
\right),
~~K=\left(%
\begin{array}{cc}
-0.5 & 1\\
\end{array}%
\right).
\end{equation*}
We adopt the two-nearest-neighbour graph to describe the  topology, i.e., $\mathcal{V}_i=\{j~|~j=i\pm1,i\pm2\}$, $i=1,2,\dots,10$. If $j\in\mathcal{V}_i$ and $j<0$ ($j>10$), then $j=j+11$  ($j=j-10$). Since the matrix $H$ has two eigenvalues on the imaginary axis of the complex plane, the network will synchronize to a stable time-varying solution determined by the initial condition. By calculating, we get $\alpha=2.9061$. We select $\delta=0.9$. Figure \ref{fig_ddt}  gives the simulation results of the network  with the distributed ETR \eqref{detre_dwell} (DDT), which shows the effectiveness of the proposed method.   In the figure, we only give the sampling time instants in the first 2 seconds for clarity. The theoretical value of $\tau^*$ is 0.0013 s. The minimum and maximum sample periods ($\tau_{min}$/$\tau_{max}$) for each node during the simulation time are given in Table 1 which shows that the actual sample periods are much larger than $\tau^*$.  

We also compared our  method with the decentralized ETR  \eqref{etr_det} (DET) proposed in \cite{Guinaldo_2013_IET}. According to Remark \ref{remak_z}, only bounded synchronization can be guaranteed with $c_0\neq 0$ in  \eqref{etr_det} (\cite{GSeyboth_auto_2013}). For this case, the advantage of our method is clear.  So here, we only compare our method with the case $c_0=0$ where asymptotic synchronization under \eqref{etr_det} can also be achieved. We select $c_1=\rho$ and $\gamma=0.30579$. During the simulation period (0 -- 18 s),  the network with DDT samples 3432 times in total, whereas the network with DET samples 212 times more (3644 times in total). 

\begin{table}[h]
Table 1. The minimum/maximum sample period 
\centering 
\begin{tabular}{c c c c c c c c c c c} 
\hline\hline 
 & Node 1 & Node 2 & Node 3 & Node 4& Node 5\\ [0.5ex] 
\hline
$\tau_{min}$& 0.0153 & 0.0114 & 0.016 & 0.0214 & 0.0188 \\
\hline
$\tau_{max}$& 0.2651 & 0.5292 &	0.6336 & 0.0817 & 0.1851 \\
\hline\hline 
 & Node 6 & Node 7 & Node 8 & Node 9 & Node 10\\ [0.5ex] 
\hline
$\tau_{min}$& 0.0046 & 0.0688& 0.0116 & 0.0117 & 0.0117 \\
\hline
$\tau_{max}$& 0.3584 & 0.2841 &	 1.4677 & 0.5347 & 0.5238\\
\hline 
\end{tabular}
\label{table_ddt_sam} 
\end{table}
\section{Conclusion}\label{sec_con}

This paper has studied asymptotic synchronization of  networks by using  distributed ETC. By using the introduced estimators,  a distributed ETR for each node has been explored, which only relies on the state of the node and states of the estimators. It has been shown that the proposed ETC synchronizes the network  asymptotically with  no Zeno behaviours.  It is worth pointing out that time-delay and data packet dropout are common phenomena which definitely affects the synchronization of networks with event-based communication. It appears that synchronization of such networks with imperfect communication is an important issue to pursue further for both theoretical interest and practical consideration. 

\begin{ack}                               
Liu's work was supported by The University of Hong Kong Research Committee Research Assistant Professor  Scheme
 and a grant from the RGC of the Hong Kong S. A. R. under  GRF through Project No. 17256516.  Cao's work was supported in part by the European Research Council (ERC-StG-307207) and the Netherlands Organization for Scientific Research (NWO-vidi-14134). De Persis's work was  partially supported by
the Dutch Organization for Scientific Research (NWO) under the auspices of the project {\it Quantized Information Control for formation Keeping} (QUICK) and by the STW Perspectief program ``Robust Design of Cyber-physical Systems" under the auspices of the project ``Cooperative Networked Systems".  Hendrickx's work was supported by the Belgian Network DYSCO (Dynamical Systems, Control, and Optimization), funded by the Interuniversity Attraction Poles Program, initiated by the Belgian Science Policy Office.
\end{ack}


\end{document}